\documentclass[11pt]{article}
\usepackage{arXiv}
\usepackage{authblk}

\usepackage[round,sort]{natbib}

\usepackage{graphicx}
\usepackage[dvipsnames,table,xcdraw]{xcolor}
\usepackage{framed}
\usepackage[normalem]{ulem}
\usepackage{amsmath}
\usepackage{amsthm,thmtools,mathtools}
\usepackage{amssymb}
\usepackage{amsfonts}
\usepackage{soul}
\usepackage{enumerate}
\usepackage{algorithmic} 
\usepackage[ruled, vlined,noend]{algorithm2e}
\usepackage{wrapfig} 
\usepackage{caption}
\usepackage{enumitem}

\usepackage[
  pagebackref,
  colorlinks=true,
  urlcolor=Blue,
  linkcolor=Blue,
  citecolor=Blue,
]{hyperref}
\usepackage[nameinlink]{cleveref}

\theoremstyle{definition}

\newtheorem{claim}{Claim}[section]

\newtheorem{definition}{Definition}[section]
\newtheorem{example}{Example}[section]

\newcolumntype{s}{>{}p{3.4cm}}

\SetCommentSty{mycommfont}
\usepackage{mathbbol}
\newcommand{\data}{\mathcal{X}^\text{rev}}
\newcommand{\datalabel}{\mathcal{Y}^\text{rev}}
\newcommand{\revx}[1]{{\mathbf{#1}}^{\text{rev}}}
\newcommand{\revy}[1]{{#1}^\text{rev}}
\renewcommand{\vec}[1]{\mathbf{#1}}
\newcommand{\datap}{\mathcal{X}^\text{rev}_{P}}
\newcommand{\datalabelp}{\mathcal{Y}^\text{rev}_{P}}
\newcommand{\datat}{\mathcal{X}^\text{rev}_{T}}
\newcommand{\datalabelt}{\mathcal{Y}^\text{rev}_{T}}
\newcommand{\datapt}{\mathcal{X}^\text{rev}_{PT}}

\newcommand{\datatt}{\mathcal{X}^\text{rev}_{TT}}
\newcommand{\datalabeltt}{\mathcal{Y}^\text{rev}_{TT}}

\usepackage{floatrow}
\usepackage[label font=bf,subrefformat=parens,labelformat=parens]{subfig}
\makeatletter
\newcommand{\HEADER}[1]{\ALC@it\underline{\textsc{#1}}\begin{ALC@g}}
\newcommand{\ENDHEADER}{\end{ALC@g}}
\makeatother

\begin{document}

\setlength{\textfloatsep}{8.5pt plus 8.5pt minus -1pt}
\setlength{\dbltextfloatsep}{7pt plus 7pt minus -1pt}

\title{The Equity Framework: Fairness Beyond Equalized Predictive Outcomes}

\author[1]{\textbf{Keziah Naggita}}
\author[2]{\textbf{J. Ceasar Aguma}}
\affil[1]{Toyota Technological Institute at Chicago,  \texttt{knaggita@ttic.edu}}
\affil[2]{University of California at Irvine, \texttt{jaguma@uci.edu}}

\date{}
\maketitle

\begin{abstract}
Machine Learning (ML) decision-making algorithms are now widely used in predictive decision-making, for example, to determine who to admit and give a loan. Their wide usage and consequential effects on individuals led the ML community to question and raise concerns on how the algorithms differently affect different people and communities. In this paper, we study fairness issues that arise when decision-makers use models (\textit{proxy models}) that deviate from the models that depict the physical and social environment in which the decisions are situated (\textit{intended models}). We also highlight the effect of  \textit{obstacles} on individual access and utilization of the models. To this end, we formulate an Equity Framework that considers equal access to the model, equal outcomes from the model, and equal utilization of the model, and consequentially achieves equity and higher social welfare than current fairness notions that aim for equality. We show how the three main aspects of the framework are connected and provide an equity scoring algorithm and questions to guide decision-makers towards equitable decision-making. We show how failure to consider access, outcome, and utilization would exacerbate proxy gaps leading to an infinite inequity loop that reinforces structural inequities through inaccurate and incomplete ground truth curation. We, therefore, recommend a more critical look at the model design and its effect on equity and a shift towards equity achieving predictive decision-making models.
\end{abstract}

\section{Introduction}\label{sec:intro}

Machine Learning (ML) algorithms are now heavily used in decision-making to determine who to hire, admit to a college, give a loan, among other decisions that have real-life consequences on people. Automated decision-making allows us to track decision-making and audit systems for fairness. However, since these algorithms use data already affected by systematic inequalities, they reinforce and sometimes exacerbate these inequalities \citep{corbettdavies2018measure,binns2018fairness}. For example, language models reveal encoded stereotypes \citep{Garg_2018,bolukbasi2016man,wordembed} which can lead to unfair decision-making \citep{De_Arteaga_2019,wordembed2}.  In addition,  \citet{ensign2017runaway} shows that predictive policing algorithms trained on biased historical data lead to more policing in those areas as more misleading crime data indicates a need for more policing. 
Consequently,  disadvantaged groups in low-income communities are more likely to be incarcerated and or denied bail \citep{machine_bias,529506}.  These issues have inspired the ML fairness community to define, design, and implement algorithmic fairness notions and techniques to tackle the introduction, propagation, and amplification of human biases in automated decision-making systems.

Prior work on fairness in ML has focused on equalizing outcomes across groups and modeled fairness violations based on the similarity between individuals and protected groups (e.g, \cite{DBLP:journals/corr/KleinbergMR16,Dwork2018IndividualFU}).  Models focused on equalized predictive outcomes sometimes acknowledge biases in historical data \citep {DBLP:journals/corr/abs-1901-10002,10.3389/fdata.2019.00013, DBLP:journals/corr/abs-1908-09635} used in decision-making.  However, historical structural inequalities like wealth, housing, and education inequities create different \textit{obstacles} for people, sometimes even within the same protected groups. These obstacles present barriers that prevent full access and utilization of the models.  The focus on only equalized predictive outcomes ignores and sometimes further excludes already marginalized\footnote{We use the terms marginalized and disadvantaged interchangeably to mean individuals facing relatively high levels of obstacles that deter them from full access and utilization of the decision-making models due to historical inequities. The term advantaged refers to individuals who don't face or face few obstacles to access or utilize the decision-making models.} groups.  

Although several ML fairness research, for example, causal and Bayesian inference (e.g., \cite{kilbertus2019sensitivity, Madras2019FairnessTC,surveyCausalBased, liu2021rawlsnet}), fair decision-making (e.g., \cite{celis2020interventions, Kleinberg2018SelectionPI, Emelianov2020OnFS}) acknowledge obstacles individuals face, the focus is mainly on changing decision-making models, for example, to qualify obstacle-refrained individuals.  We, however, argue that this only helps in the short and not the long term because unalleviated obstacles individuals face in accessing the model resurface in the model utilization, in which decision-makers evaluate individuals on how well they utilize the model as a form of feedback to the decision-makers. In addition, since decision-makers use models (\textit{proxy models}) that deviate from the social and physical environment in which the decisions are situated (\textit{intended models}), inequality in utilization increases as individuals are likely to face unforeseen obstacles unalleviated during decision-making. Additionally, the discrepancy increases the chances of curating incomplete, inaccurate, and skewed ground truth. 

\example{(Pretrial/Bail Assessment Model)}\label{ex:bail}
Most bail models use a model function like Public Safety Assessment (PSA) algorithm \citep{harvardlawreview} to determine who to give or deny bail.  The function uses features: ``age at current arrest'', ``current charges'', ``pending charges'', ``prior misdemeanor conviction'', ``prior felony conviction'', ``prior violent conviction'', ``prior failure to appear in past two years'',  ``record of failing to appear'',  and ``record sentences to incarceration''.  An individual facing obstacles like lack of good representation might not have someone accurately layout/dispute current criminal charges and criminal history,  for example, when they add the defendant's parents' charges to their sentences \citep{recidivism}. Another might live in a neighborhood with higher chances of rearrest, thus leading to an accumulative crime history and high record failure to reappear.   For example, when people are released, some have to go to halfway houses or get probation or parole officer (PPO) residing in the same place they got convicted, and or in poor neighborhoods surrounded by crime \citep{halfwayprison_legal,newyork,halfwayhouses}.  Going back to the same environment increases chances of falling into the same habits that got them in trouble and chances of rearrest as the place is more likely to be frequently patrolled \citep{ensign2017runaway}. 
These obstacles disproportionately affect some groups than others, and thus leading to a high level of unequal access to the bail model. Therefore, decision-makers should ask themselves: \textit{with this choice of predictive features: current charges, criminal history, and record of failing to appear, among others who is more(less) likely to access this model? Who is more likely to be gain(lose) by it? }

In addition, to determine whether someone is a true or false positive, the bail model relies on whether someone appears for their trial in court.  However, the proxy model did not capture features that might determine whether someone reappears, for example, ``support system'', ``financial stability'', ``job lineup'', among others.  Therefore, an intended model with a model function ``likely to reappear in court'' and those features would have been a better predictor. Using the intended model function and features that depict the social and physical environment that decisions are situated helps highlight obstacles that might lead to individuals not fully utilizing the model and appearing as false positive.  For example, insufficient support from PPOs due to heavy case-loads \citep{Okonofuae2018036118}, non-flexible jobs with no time off, or even being too busy to remember the court date \citep{harvardlawreview,lawnotifications} might be highlighted and alleviated.

\example{(Loan worthiness)}\label{ex:p2p_lending} 
Similar to example \ref{ex:bail}, determining who to grant a loan is an example of a delayed evaluation model. Delayed evaluation models, unlike immediate ones, require decisions to take form in the environment before the decision-maker can determine if the predictions are true or false positives. In their proxy model used to determine who to give/deny a loan, \citet{Turiel2019P2PLA} used model functions like logistic regression on proxy features ``debt to income ratio'', ``employment length'', ``loan amount'', and ``loan purpose''. However, to evaluate their predictions to know if they are true or false positive, they used another model (intended model) to determine who paid/defaulted on the loan. The intended model had model function support vector machine and was trained on intended features e.g. ``loan amount'', ``term'', instalment'', ``employment length'', ``income verification status'', ``homeownership'', ``FICO score'', and ``public records.  

While it's common in delayed evaluation models to have the proxy model different from the intended model, our work shows that the consequences on the different individuals are different, consequential, and possibly long-term. Ignorance of the discrepancy between the proxy and intended models prolongs unfairness, leads to incomplete ground truth curation, and further excludes already marginalized populations. Additionally, this discrepancy renders models that don't alleviate obstacles individuals face but rather employ other interventions like changing the performance/accuracy metrics, changing thresholds/model functions (e.g., \citep{celis2020interventions,Kleinberg2018SelectionPI,Emelianov2020OnFS}) ineffective. The reason is while obstacle-refrained individuals might receive a desirable outcome (e.g., granted a bail), the obstacles they faced in accessing the model remain. These obstacles resurface in model utilization, in addition to utilization obstacles, thus increasing the individual's likelihood for unequal utilization. Therefore, in the evaluation phase, the individual appears as a false positive, that is to say, fails to reappear in court. The decision-maker who is oblivious to the discrepancy between the model they used to make decisions and the one they use to evaluate the decisions uses this label to curate new ground truth, thus reinforcing biases and creating a cycle of inequities.

Therefore, while lots of work has gone into understanding decision-making systems and making them fairer, for example, the push towards no money bail \citep{bail_reform_ca,bail_reform}, our work provides a deeper understanding of how different the consequentiality of delayed evaluation models are from immediate ones.  This work attempts to show how the difference between proxy models and the models that depict the physical and social environments that the decisions are situated differently affects different people and factors tremendously in the curation of future proxy model ground truth. When ignored and misunderstood, the proxy gap makes it harder to diagnose problems accurately, find equitable solutions, and implement policies that help the most disadvantaged. 

Additionally, we expose the pitfalls of achieving equalized outcomes without access,  changing decision-making models ---features, labels, and model functions-- without alleviating obstacles individuals face, and using proxy models that deviate far from intended models. The Equity Framework motivated by \citet{almond} provides decision-makers with a path to follow with checkpoints for equal access, outcomes, and utilization to ensure their models are equitable, achieve high social welfare, and provide long-term fairness. We believe our work will help in the insurance, evaluation, and auditing for fairness in predictive decision-making models.

\paragraph{Our Contributions are summarized as follows:}
\begin{enumerate}
\item[\textbf{1.}] We show how obstacles, a result of historical inequities, create the implicit and explicit barriers that deter individuals from accessing and fully utilizing models (sec.\ref{sec:main_prel} and sec.\ref{sec:framework}). With an example, we prove that it's possible to achieve equal outcomes without equal access, which further marginalizes already marginalized groups, thus reinforcing structural inequities.  

\item[\textbf{2.}] We differentiate immediate evaluation models from delayed ones and show how the proxy gap between proxy and intended models negatively affects model utilization, future ground truth curation, and long-term fairness. We supplement our mathematical formulations with examples (sec.\ref{sec:main_prel} and sec.\ref{sec:framework}).

\item[\textbf{3.}] We define and connect all the three aspects of the Equity Framework, equal access, equal outcomes, and equal utilization. We provide an algorithm (sec.\ref{sec:equity_score}) and a set of questions to guide decision-makers towards equitable decision-making (sec.\ref{sec:equity_score}) . Using student admission as a case study, we empirically show the flow of the Equity Framework and illustrate challenges that might be faced in practice when trying to follow the Equity Framework (sec.\ref{sec:case_study}). 
\end{enumerate}

A key insight from our work is that equal outcomes as a fairness notion is not enough to address disparities arising from the discrepancy between the proxy and intended models in delayed evaluation systems and different obstacles individuals face to access and utilize models.
Generally, social welfare achieved from the Equity Framework surpasses the currently considered fair outcome-based predictive models.  We, therefore, argue for a shift in fairness literature to consider fairness beyond predictive model outcomes. We hope our work pivots the discussion on fair predictive decision-making from equality to equity and becomes a yardstick for decision-makers to check their models for equitable decision-making. 

\paragraph{Related Work} \label{sec:related_works}
To achieve equalized fair decision-making, researchers have proposed several metrics under the umbrella of group fairness and individual fairness. Group fairness ensures positive or negative parity in the treatment of protected groups \citep{DBLP:journals/corr/Chouldechova17,DBLP:journals/corr/KleinbergMR16}, and individual fairness ensures similar treatment of similar individuals per the decision-making model \citep{DworkFairness}.  However, historical structural inequalities like wealth, housing, and education inequities create \textit{obstacles} which present barriers that prevent full utilization and access to models.  Focusing on only equalized predictive outcomes ignores and sometimes further marginalizes already historically excluded groups. 

Some researchers have highlighted the issues of access disparities among populations. For example, strategic classification (e.g, \citep{miller2020strategic,hardt2015strategic,dong2017strategic,Hu_2019,ahmadi2021strategic,milli,kleinbergstrategic,frankel2021improving}) shows how people's reactions to the decision-making model highlights the differences in access through costs people incur. Although our framing of obstacles is quite analogous to budget framing in strategic classification, we formulate obstacles more generally, and the alleviation of obstacles strictly makes things better, that is to say, the obstacle-free feature values, $\vec{z}$ dominate the obstacle-refrained feature values $\vec{x}$ and the obstacle-free label $y'$ is not necessarily equal to the obstacle-refrained label, $y$. Relatedly, another body of work that highlights obstacles individuals face is causal and Bayesian inference
(e.g., \citep{kilbertus2019sensitivity, Madras2019FairnessTC,surveyCausalBased, liu2021rawlsnet}). However, instead of ensuring equal access, the focus is mainly on redefining decision-making models, by, for example, changing accuracy metrics, weights of different features, or features used, among other interventions to qualify obstacle-refrained individuals.  In our work, we show that while redefining decision-making models might grant the obstacle-refrained individuals a foot in the door, these obstacles resurface in the evaluation phase, leading to further marginalization of the disadvantaged groups. 

Just like one of the main aspects of the Equity Framework, \citet{rawls_book} asserts that to achieve fairness, individuals with the same level of talent, ability, and willingness to use those gifts should have a fair chance at reaching desirable positions without the obstacles of their social class and background.  While the concept of equity in machine learning is relatively new and mostly foreign, equity has been a goal for many reforms in justice \citep{nicholson2009exploring}, health \citep{almond}, and education \citep{gorard2004international} systems.  Recently, however, the concept has started taking shape and has drawn a few researchers to it. For example, \citet{red-max} take the causal perspective to show that while predictive parity might achieve fairness across groups, it might perpetuate inequality within groups and legitimize the status quo. In doing so, \citet{red-max} bring forward new questions on the power distributions between and within groups as it relates to decision making systems. \citet{suresh-guttang}'s major contribution is a framework for identifying biases that arise due to historical content within which the ML development pipeline is situated. Similar to ours, their (\citet{suresh-guttang}) framework is a  push towards more equitable decision-making highlights the downstream harms and how they manifest in model building, evaluation, and deployment processes. More closely related to our work is \citet{DBLP:journals/corr/abs-2005-07293} who attempts to formalize equity by equalizing the sum of historical plus future outcomes of one group to another to compensate for observed historical biases in the data. However, unlike \citet{DBLP:journals/corr/abs-2005-07293}, we focus on ensuring obstacles caused by historical biases are alleviated to achieve equal access to the model. In our work, we argue that, if not alleviated, obstacles resurface in utilization, thus widening the inequity gap. Therefore, the basis of the Equity Framework is motivated by \citet{almond} who defines equity as insurance that barriers that could deter some people from having full access to the available resources are alleviated.  

Apart from forgetting to check for and ensure access, decision-makers more often than not use proxy models instead of intended decision-making models. This is mainly due to bounded rationality  \citep{Simon1990}, past experiences, biased view of the world \citep{pastfuture, Stanovich2008OnTR, Bruin2007IndividualDI,decisiontovote}, difficulty articulating what they want to measure \citep{counts_counted,10.2307/2631637}, among others. In this paper,\textit{ we define a proxy model as the model decision-makers use in making decisions}. We define the \textit{intended model as the model that depicts the environment in which made decisions take form}. The ignorance of the discrepancy between proxy and intended models increases unequal utilization and curation of biased ground truth. 

Several researchers have studied the origin, diversity, and accuracy of ground truth data used in decision-making and its effect on predictions and different populations. For example, \citet{gender_shade} showed that bias in ground truth was one of the leading causes of higher error rates on dark-skinned women than other protected groups.  To better understand the origin of ground truth, \citet{datset_sheets} formulated a way for researchers to ensure responsible data collection and documentation. \citet{Jacobs_2021, doi:10.1177/20539517211013569,sogaard-etal-2014-selection,app10114014} and \citet{10.1117/1.JRS.13.034522} have contested and questioned the validity of ground truth, and \citet{aka2021measuring} proposed models of measuring bias in classification independent of ground truth. Our work adds to this body of work to show how the gap between proxy and intended model leads to faulty, inaccurate, and incomplete ground truth curation. We also provide a way to measure these proxy gaps and present a novel idea of an Equity Framework whose aim is to ensure equal model access, outcome, and utilization and facilitate efficient auditing of deployed models for equity.

\section{Preliminaries}\label{sec:main_prel}
To understand the main attributes of the Equity Framework; access, outcomes, and utilization of the model, we provide an overview on obstacles and proxy gaps.

\subsection{Obstacles} \label{sec:intro_obstacles}
In this paper, we define \textit{obstacles as the implicit and explicit barriers that deter individuals from effectively interacting with the decision-making model.} We assume that the decision-maker uses features as inputs to their models, and barriers implicitly and explicitly affect these inputs. An individual with feature representation $\vec{z} \in \mathbb{R}^d$, will instead have feature representation $\vec{x} \in \mathbb{R}^d$ due to obstacles faced when interacting with the decision-making model.  For example, a work-study student with knowledge of test scores factoring greatly in their performance is less likely to prepare adequately or attend enough discussions and will have feature representation $\vec{x}$ instead of $\vec{z}$ due to lack of ample time. The  student doesn't show up to the model as they could have, had their obstacles been alleviated.  We say that $\vec{z}$ \textbf{dominates} $\vec{x}$, that is $\vec{z} \succ \vec{x}$ because, $$\forall i \in [d], \vec{z}_i \geq \vec{x}_i \hspace{4px} \text{and} \hspace{4px} \exists i \in [d],  \vec{z}_i > \vec{x}_i $$
Mathematically, we represent obstacles as: $$\mathcal{O}(\vec{x}, \vec{z}) = \langle\alpha,\vec{z}-\vec{x}\rangle \in \mathbb{R}_{\geq 0}$$ where $\alpha \in \mathbb{R}^d$ is how much the implicit and explicit barriers constrain an individual from effectively interacting with the decision-making model.
To improve an individual's access to the model means alleviating these obstacles, $\mathcal{O}(\vec{x}, \vec{z})$, such that the likelihood, $P(\mathcal{Y} | \mathcal{X})$ of effective interaction with the model increases $P(\mathcal{Y} = 1 | \mathcal{X} = \vec{z}) \geq P(\mathcal{Y} = 1 | \mathcal{X} = \vec{x})$. 

The assumption that alleviating the obstacles leads to $\vec{z}$ dominating $\vec{x}$ and improves the likelihood of desirable prediction might not capture all the complexities of inequities effects on individuals and relationships of features to labels.  Future works could explore a more generalized formulation. Additionally, within any decision-making system, there could be several obstacles to the access and utilization of the model, most specific to the decision to be made. We assume a finite number of obstacles that directly affect access to and utilization of any given decision-making system. We imagine to be obstacles a decision-maker has full knowledge of and can create policies to alleviate said obstacles.
\subsection{Proxy gap} \label{sec:intro_gaps}
Evaluation of some ML predictive decision-making algorithms is immediate and straightforward, while in others, evaluation is not immediate and most times done with a separate model. For example, after predicting a cat from the image, the ML decision-maker only has to compare the picture with the prediction to know if it's a true positive or false positive prediction. However, for other tasks like predicting loan worthiness, the evaluation is not immediate. The decision-maker only knows who is a true positive or false positive after the decisions take form in the social and physical environment. Usually, the models used to make decisions (proxy models) and those used in evaluations (intended models) are different.
In this work, we define the \textit{proxy gap as the discrepancy between the model decision-makers use in making decisions (proxy model) and the model that depicts the environment in which decisions take form (intended model)}.  We measure this gap as the discrepancy between proxy features and intended features, proxy label function\footnote{For language simplicity, we sometimes use ``label'' instead of ``label function'' or ``model function''. The term (decision-making) model defines a decision-making system that takes in features and obstacles, has a set of policies to alleviate the obstacles individuals face to access and utilize the model, and has a class of label functions that output decisions.} and the intended label function,  and access obstacles and utilization obstacles. We then explore how the proxy gap affects model utilization and future ground truth curation.  In examples \ref{ex:bail} and \ref{ex:p2p_lending}, we described how the proxy model might be different from the intended model.

\section{Attributes of the Equity Framework}\label{sec:framework}

In this section we give a detailed description of the attributes of the Equity Framework ---equal access, outcome, and utilization---, their formulations and how they are related.

\subsection{Equal Access} \label{subsec:equal_access}
If a decision-making model, $f$, is such that individuals face obstacles, $\mathcal{O}_{f}(\vec{x}, \vec{z})$, that hinder their effective interaction with the model, it becomes an equal access model when obstacles all individuals face are alleviated. The model's policy, $\Phi$, ensures that the obstacles each of the individual faces are reduced to $0$. Therefore, $\Phi$ is defined as, $$\Phi(\mathcal{O}_{f}(\vec{x}, \vec{z}), \delta) = \max\{\mathcal{O}_{f}(\vec{x}, \vec{z}) - \delta, 0\}$$ Here $\delta \geq 0$ is the decision-maker's resource budget for alleviating obstacles individuals face, where $\mathcal{O}_{f}(\vec{x}, \vec{z}) \ll \delta$ implies a surplus of resources.

\begin{figure}[htp!] 
    \centering
    \includegraphics[width=6cm]{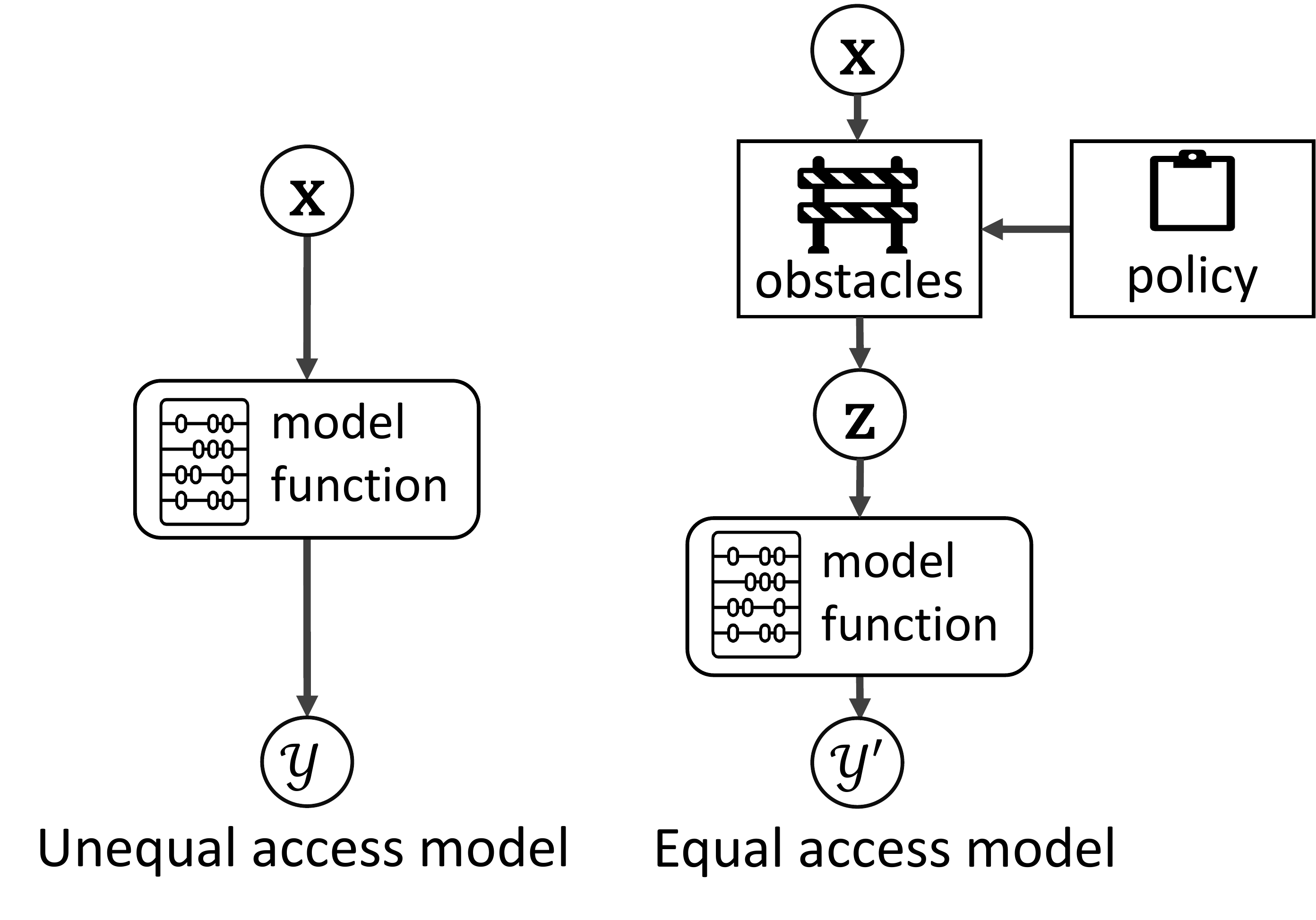}
    \vspace{-3mm}
    \caption{On the left, the model doesn't check for or alleviate obstacles, and, on the right, is an equal access model which uses a policy to alleviate obstacles individuals face when interacting with the model.}
    \label{fig:access}
\end{figure}
\raggedbottom

\definition{(Model access)} Given a model, $f$, assume each individual, $i$, faces obstacles $\mathcal{O}_{f,i}$  to access the model. An individual achieves full access to the model if they either face no obstacles to access the model, $\mathcal{O}_{f,i}(\vec{x}_{f,i}, \vec{z}_{f,i}) = 0$ or if the model checks for and alleviates all the obstacles an individual, $i$ ,faces, $\Phi_{f}(\mathcal{O}_{f}(\vec{x}_{f,i}, \vec{z}_{f,i}), \delta) = 0$.

Model access based on whether individuals have full access to the model is defined as, 
$$\displaystyle \Psi(f) = \frac{\sum_{i=1}^{n}\mathbb{1}_{\{\Phi_{f}(\mathcal{O}_{f,i}(\vec{x}_{f,i}, \vec{z}_{f,i}), \delta) = 0\}}}{n}$$ 
Therefore, if $\Psi(f) = 1$, the model, $f$, has \textbf{equal access}.  Additionally, during decision-making, the individual  (removed $i$ to avoid notation clutter) reveals themselves as $\revx{x}_{f}$ and the decision-maker sees them as $\revy{y}_{f}$, (fig.\ref{fig:access}) where, 
\begin{equation*}
\begin{split}
(\revx{x}_{f}, \revy{y}_{f}) =  \left\{
    \begin{array}{l}
    (\vec{z}_f, y^{'}_{f}), \ \text{if} \ \mathcal{O}_f(\vec{x}_{f}, \vec{z}_{f})= 0 \ \text{or}  \ \Phi_{f}(\mathcal{O}_{f}(\vec{x}_{f}, \vec{z}_{f}), \delta) = 0\\
    (\vec{x}_f, y_{f}), \ \text{o/w}
    \end{array}
  \right.
\end{split}
\end{equation*}

\example{} Individuals in different domains face different obstacles to access the models in those domains. For example, participant 8 in a subject study conducted by \citet{parttimestudy} to assess the impact of term-time employment on students, said ``I get less time to focus on my assignments and to do my reading and prepare for my lectures.'' Because they simultaneously work and study, they don't perform as they should have if they had a scholarship or financial aid (policy). Another example is that of high child care demands decreasing employee performance \citet{childcaredemands}. Perhaps if the workplace provided childcare services, the employee could have been more effective at work. 

\subsection{Equal Outcomes} \label{subsec:equal_outcomes}

Fairness in predictive outcomes has been well covered in the algorithmic fairness literature. In this paper we adopt equalized odds (EO) \citep{DBLP:journals/corr/HardtPS16} fairness metric as a measure of model outcomes. Other parity measures for example, demographic parity \citep{feldman2015certifying}, accuracy parity \citep{doi:10.1177/0049124118782533} can be used. 

Consider a binary classification setting and population distribution, $\mathcal{D}$ of size $n$. Each individual with protected group membership $grp \in \mathcal{G}  = \{0, 1\}$ reveals features $\revx{x} \in \data \subseteq \mathbb{R}^{d}$, and has a label $\revy{y} \in \datalabel = \{0, 1\}$.  To make a prediction, a decision maker employs a model function, $h : \data \to \datalabel$ that generalizes with minimal accuracy loss, $\min_{h \in \mathcal{H}} L(h, \mathcal{D})$ where $\mathcal{H}$ is the set of all possible model functions and $L(h, \mathcal{D})$ is the loss function. Throughout this paper, for simplification and WLOG, we assume that there is one protected feature. We define EO violation as $[P(h(\revx{x}) = 1|\revy{y}=1, \ grp = 0) - P(h(\revx{x}) = 1|\revy{y}=1, \ grp = 1)] + [P(h(\revx{x}) = 1|\revy{y}=0, \ grp = 0) - P(h(\revx{x}) = 1|\revy{y}=0, \ grp = 1)]$

\definition{(Model outcomes)} We define the outcomes of the model, $f$ as $\Omega(f) = $ EO violation. If the EO violation $= 0$, the model outcomes,  $\Omega(f) = 1$, indicating that the model achieves \textbf{equal outcomes}. 

\example{} An example of a model with unequal outcomes is the Propublica model in which the model function in the Prater-Borden case \citep{mach} was ``particularly likely to falsely flag black defendants as future criminals, wrongly labeling them this way at almost twice the rate as white defendants'' \citep{mach}. 
 
\subsubsection{Model outcomes and access} \label{sec:outcomes_access}
\begin{figure}[htp!]
    \centering
    \includegraphics[width=10cm]{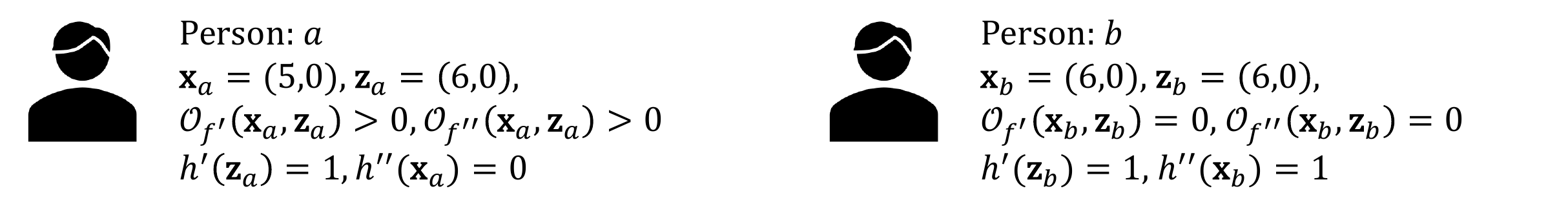}
    \caption{When interacting with models $f'$ and $f''$, individual $a$ on the left is faced with obstacles, and individual $b$ on the right is not faced with obstacles.}
    \label{fig:income_acc}
\end{figure}

To illustrate how model access relates to model outcomes, let's assume we have two individuals, $a$ and $b$, and two models: an equal access model, $f',$ and an unequal access model, $f''$, whose label functions respectively are,  $h'(\revx{x}) = 1, \ \text{if $\|\revx{x}\|\geq 5.5$}$, and $0 \ \text{otherwise} $ and $h''(\revx{x}) = 1, \ \text{if $\|\revx{x}\|\geq 6$}$, and $0 \ \text{otherwise}.$ Person $a$ faced with obstacles to access $f'$ and $f''$ and $b$ not faced with obstacles to access both models (see fig.\ref{fig:income_acc}). 

\raggedbottom

With the unequal access model, $a$ reveals $\vec{x}_a = (5,0)$, and is predicted as $h''(\vec{x}_a) = 0$, and with the equal access model, $a$ reveals $\vec{z}_a = (6,0)$ and $h'(\vec{z}_a) = 1$. On the other hand, $b$ doesn't face obstacles to present themselves to both models, and their feature values $\vec{x}_b = \vec{z}_b = (6,0)$. Person $b$ is predicted as $h'(\vec{z}_b) = 1$ and $h''(\vec{x}_b) = 1$ (fig.\ref{fig:income_acc}). From the example, since both model functions rely on information they receive, they achieve  equal outcomes even though  $f''$ doesn't ensure equal access.

We can therefore see that for any equal access model $f'$, there is an unequal access and equal outcome model $f''$, such that; $\Psi(f'') < \Psi(f')$  and $\Omega(f'') = \Omega(f').$ Therefore, with equality fairness notions like equalized odds, the model function because it relies on received data can appear to achieve fairness when it's actually exacerbating inequity through unequal access. This setting further marginalizes disadvantaged groups, and the problem is decision-makers don't even realize the gravity of the error. We, therefore, urge the  ML fairness community to check for equal access before equal outcomes.

\subsection{Equal Utilization} \label{subsec:equal_util}

\paragraph{Setting} \label{sec:model_setting}
Assume a given proxy model, $P$, with a model function, $h_{P} \in \mathcal{H}_{P}: \datap \to \datalabelp$, is trained on proxy features $\revx{x}_{P} \in \datap \subseteq \mathbb{R}^{d}$ and proxy target labels $y^{\text{rev}}_{P} \in \datalabelp = \{0,1\}$. Assume the trained model function achieves maximal accuracy with minimal loss, $\min_{h_{P} \in \mathcal{H}_{P}} L(h_{P}, \mathcal{D}_{P})$, where $L(\cdot)$ is the loss function.  The best trained proxy model function, $h_{P}$, is then tested in the wild on new arriving individuals. The arriving individuals have similar proxy feature variables as those considered in training, $\datapt$. Individuals are faced by access obstacles $\mathcal{O}_P: \mathcal{X}_{P}\times\mathcal{X}_{P} \to \mathbb{R}_{\geq 0}$, and depending on whether the proxy model has equal access or not determines $\revx{x}_{PT} \in \datapt$ as described in section \ref{subsec:equal_access}. The trained proxy model, $h_{P}$, then predicts them as $y_{PT} \in \mathcal{Y}^{\text{rev}}_{PT}$.

\begin{figure}[!ht]
    \centering
    \includegraphics[width=0.8\linewidth]{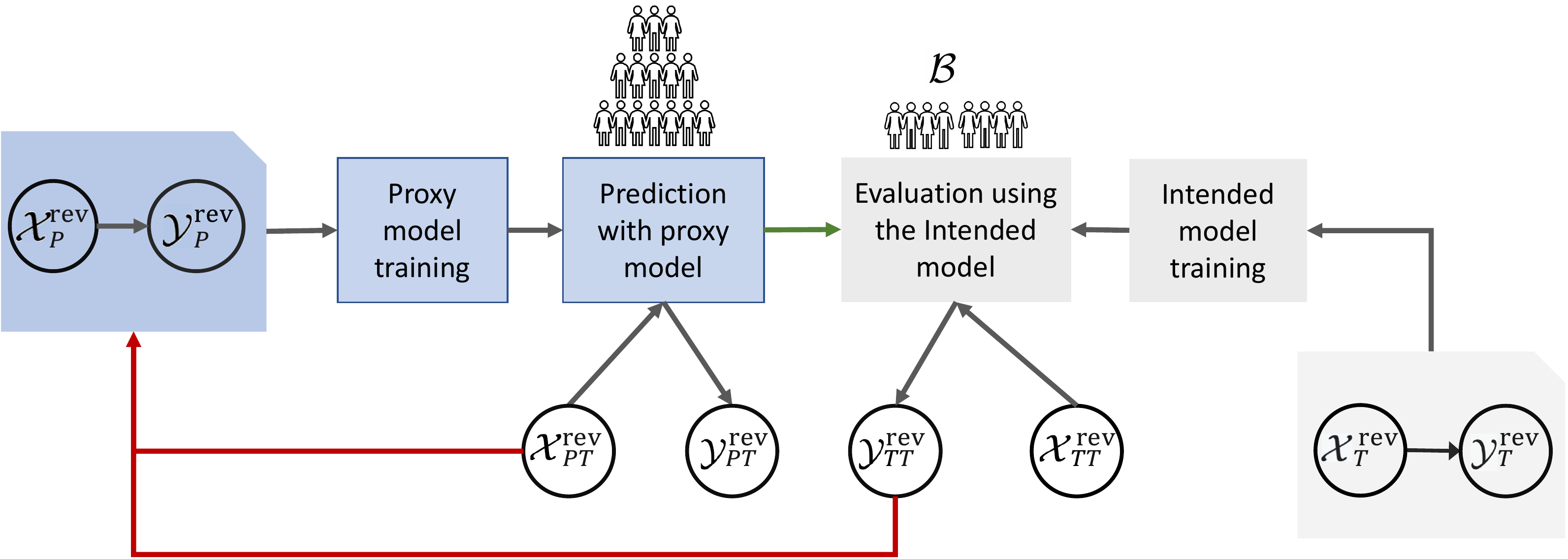}
    \caption{The explicit proxy decision-making model (light blue) is trained on ground truth and tested in the wild. All positively classified individuals from the proxy model are evaluated for utilization using the implicit intended model  (light grey).  Decision-makers ignorant of the proxy gap, use the utilization feedback $\mathcal{Y}^{\text{rev}}_{TT}$ and proxy features $\mathcal{X}^{\text{rev}}_{PT}$ to curate ground truth (red lines).}
    \label{fig:equity}
\end{figure}
\raggedbottom

Let all the individual qualified by the proxy model be in $\mathcal{B}$, such that, $\forall i \in \{1,\cdots,m\}, \ y_{PT,i}^{\text{rev}} = 1$, where $m = |\mathcal{B}|.$ To evaluate full utilization, we assume a pre-existing intended model, $T$ with trained model function $h_{T} \in \mathcal{H}_{T}: \datat \to \datalabelt$, trained on intended features  $\datat \subseteq \mathbb{R}^{D}$ and intended target labels $\datalabelt = \{0,1\}$. 
To evaluate how well individuals in $\mathcal{B}$ utilize the model, decision-makers have to wait for the decisions to take form in the physical and social environment. In attempting to utilize the model, individuals might face utilization obstacles  $\mathcal{O}_T: \mathcal{X}_{T}\times\mathcal{X}_{T} \to \mathbb{R}_{\geq 0}$. Depending on whether the intended model alleviates utilization obstacles or not determines how individuals in $\mathcal{B}$ reveal themselves to the intended model as described in section \ref{subsec:equal_access}. Individuals in $\mathcal{B}$ reveal themselves as $\revx{x}_{TT} \in \datatt$ whose variables are similar to $\datat$ and the intended model function then predicts them as $y_{TT} \in  \datalabeltt$.

\definition{(Model utilization)} A proxy model qualified individual, $ i \in [m]$, achieves full model utilization if and only if they remain qualified in the physical and social environment in which the model takes form. That is to say, $y_{PT,i}^{\text{rev}} = y_{TT, i}^{\text{rev}} = 1$. Model utilization $\zeta(f)$,  is then defined as $$\zeta(f) = \frac{\sum_{i=1}^{m}\mathbb{1}_{\{y_{PT,i}^{\text{rev}} = y_{TT, i}^{\text{rev}}\}}}{m}$$ 
A model, $f$, therefore has \textbf{equal utilization} if $\zeta(f)= 1$. 

\example{} First, \citet{Turiel2019P2PLA} separated loan-worthy from non-loan worth individuals. They then used the intended model to determine if loan-worthy individuals $\forall i \in \{1,\cdots,m\}, \ y_{PT,i}^{\text{rev}} = 1$ were true positive (paid back the loan) or false positive (defaulted on the loan). Analysis showed that $15\%-20\%$ didn't fully utilize the model, that is to say, they defaulted on the loan and their $y_{TT}^{\text{rev}}$ was $0$.

\paragraph{Ground truth curation} Other than higher chances of unequal utilization, one of the biggest spillover effects of ignorance of the discrepancy between intended model and proxy model in delayed evaluation models is incorrect and incomplete ground truth curation. Decision-makers oblivious to the intended model and it's influence on how individuals predicted positive by the proxy model utilize the model, use the proxy model features $\datapt$ and intended model labels $\datalabeltt$ to curate ground truth as shown with red lines in figure \ref{fig:equity}. Because of this, the next trained proxy model is trained on incomplete ground truth that falsely portrays obstacle restrained individuals, doesn't capture true feedback from the model deployment in the wild, and consequently leads to inequities.

\subsubsection{Feature, label and obstacle proxy gaps} \label{sec:proxy_gaps}
For readability, we eliminate the superscript $\text{rev}$ from the definitions of the proxy gaps. 

\paragraph{Feature proxy gap} The feature proxy gap, $\Gamma_{X} (\mathcal{X}_{P}, \mathcal{X}_{T})  \in \{0,1\}^{D}$, where $D$ is the number of intended features, defines how different the proxy model features variables are from the intended model feature variables. 

\begin{equation*}
\begin{split}
\text{Therefore}, \ \forall i \in [D], \ \Gamma_{X}(\mathcal{X}_{P}, \mathcal{X}_{T})_{i} =  \left\{
    \begin{array}{l}
    0, \ \text{if for $i$,  $\exists j: \mathcal{X}_{T}[i] \equiv \mathcal{X}_{P}[j]$}\\
    1, \ \text{o/w}
    \end{array}
  \right.
\end{split}
\end{equation*}
When $\Gamma_{X} (\mathcal{X}_{P}, \mathcal{X}_{T}) = (0,\cdots, 0) \in \mathbb{R}^{D} = \vec{0}_D$, the feature proxy gap is non-existent. 

\example{(High feature proxy gap)} 
Assume the proxy model function uses features $\{a, b, c\}$ and intended model function uses featured $\{u,v,w\}$. Now, if feature $u$ is such that none of the features $\{a,b,c\}$ is exactly similar to $u$, then $\Gamma_{X} (\mathcal{X}_{P}, \mathcal{X}_{T})_1 = 1$. Similarly, if the same applies to features $v$ and $w$, then the feature proxy gap $\Gamma_{X} (\mathcal{X}_{P}, \mathcal{X}_{T}) = (1,1,1)$, indicating that none of the features used in the proxy model is exactly similar to those in the intended model.  For example, assume a bail model determining whether or not defendant reappears in court uses proxy model features ''\textit{criminal history}'', ``\textit{current crime}'', and ``\textit{age at arrest}'' and the intended model features ``\textit{job}'', ``\textit{support system}'', ``\textit{financial stability}'' will have a feature proxy gap $\Gamma_{X} (\mathcal{X}_{P}, \mathcal{X}_{T}) = (1,1,1)$.

\paragraph{Label proxy gap} To compute the label proxy gap we first compute the feature importance scores for each model's features. Assume $h_{P}$ and $h_{T}$ are the same class of functions, for example both are logistic regression models. Let $\omega_{P}$ be the feature importance scores of proxy model function $h_{P}$, and let $\omega_{T}$ be the feature importance scores of intended model function $h_{T}$. The label proxy gap is then $\Gamma_{L}(h_{P}, h_{T})  \in \mathbb{R}^{D}$, where $D$ is the number of intended features. 

\begin{equation*}
\begin{split}
\text{Therefore}, \ \forall i \in [D], \  \Gamma_{L}(h_{P}, h_{T})_{i} =    \left\{
    \begin{array}{l}
    \omega_{T,i} - \omega_{P,j}, \ \text{if for $i$, $\exists j$} : \mathcal{X}_{T}[i] \equiv \mathcal{X}_{P}[j] \ \text{and} \ \text{sign}(\omega_{T,i}) = \text{sign}(\omega_{P,j})\\
    \omega_{T,i}, \ \text{o/w}
    \end{array}
  \right.
 \end{split}
\end{equation*}
When $\Gamma_{L}(h_{P}, h_{T}) =  (0,\cdots, 0) \in \mathbb{R}^{D} = \vec{0}_{D}$, then the label proxy gap is non-existent. While the difference between feature importance scores might be the best indication of the discrepancy of the value of the different features to the proxy and intended models, our formulation might not generalize well to all model settings.

\claim{If $\Gamma_{X} (\mathcal{X}_{P}, \mathcal{X}_{T}) \neq \vec{0}_D$, then  $\Gamma_{L}(h_{P}, h_{T}) \neq \vec{0}_D$ and $\Gamma_{X} (\mathcal{X}_{P}, \mathcal{X}_{T}) = \vec{0}_D$ does not imply $\Gamma_{L}(h_{P}, h_{T}) = \vec{0}_D$.}
\proof From the definitions of the feature proxy gap, $\Gamma_{X} (\mathcal{X}_{P}, \mathcal{X}_{T})$, and label proxy gap, $\Gamma_{L}(h_{P}, h_{T})$, we can see that if the $\Gamma_{X} (\mathcal{X}_{P}, \mathcal{X}_{T}) \neq \vec{0}_D$, then the label proxy gap,  $\Gamma_{L}(h_{P}, h_{T}) \neq \vec{0}_D$ since $\exists i,$  such that $\Gamma_{L}(h_{P}, h_{T})_{i} = \omega_{T,i}$. 

If the feature proxy gap, $\Gamma_{X} (\mathcal{X}_{P}, \mathcal{X}_{T}) = \vec{0}_D$, then assuming the feature importance scores of the proxy model  $\omega_{P}$ and intended model  $\omega_{T}$ are equal, then the label proxy label $\Gamma_{L}(h_{P}, h_{T}) = \vec{0}_D$. However, if the feature importance scores are not equal, then $\forall i \in [D], \ \Gamma_{L}(h_{P}, h_{T})_{i} =  \omega_{T,i} - \omega_{P,j}$.

\paragraph{Obstacle gap} Access obstacles are those alleviated in accessing the proxy model, and utilization obstacles are those alleviated in utilizing the intended model. The obstacle gap looks at how different utilization obstacles are from the access obstacles. 
The bigger the feature proxy gap, the higher the obstacle gap. Additionally, since determining model utilization is implicit, utilization obstacles are less likely to be alleviated as the decision-maker might not be aware of the extent of the effect of their existence. Therefore an increase in the obstacle gap likely increases unequal utilization.

\subsubsection{Model outcomes and utilization} \label{sec:outcomes_util}
To determine whether the individuals the proxy model classified positively are true or false positives requires getting feedback on utilization, a result of evaluation using the intended model (light grey on fig. \ref{fig:equity}).  When the proxy model doesn't achieve equal outcomes and disproportionately misclassifies individuals from a given protected group as false negative, it denies them a chance to be evaluated for utilization and the decision-maker doesn't get proper feedback. This is because to assess whether one is truly positive or falsely positive, the decision-maker only does this assessment (evaluation by the intended model) on those classified positive by the proxy model. 

On the other hand, proxy model false positive errors grant unqualified individuals a chance to be evaluated for full model utilization in which two consequences unfold. (1) If members of a given marginalized group face unalleviated utilization obstacles, they are less likely to have full model utilization. The decision-maker will then have negative associations with the group, which they treat as objective feedback. In addition, since the decision-maker uses the feedback ($\datalabeltt$) in ground truth curation, these half truths become part of the future decision-making. (2) However, if a member from a given advantaged group predicted falsely positive by the proxy model ends up facing no obstacles in presenting themselves to the intended model, they will have full utilization. The decision-maker will further associate positive feedback to this individual and group and feed this back into the model. Therefore the advantaged will appear more qualified even when the system errors on them, and disadvantaged individuals will look more unqualified, which further marginalizes disadvantaged groups in future decision-making. 

\subsubsection{Proxy gaps and equal utilization} \label{sec:gaps_util}
Proxy gaps increase unequal model utilization. For example, assume a judge, instead of using the intended model ``likelihood to reappear in court'', to make bail decision, uses the proxy model, ``discipline to reappear in court.'' Then the proxy model positively classified (granted bail) offenders faced by utilization obstacles affecting features like transport/airfare might appear as false positives during evaluation phase. Had the court used the intended model in the first place, the obstacles would potentially have been detected, hopefully alleviated and led to full model utilization.

Additionally, if the proxy gap is large, ground truth curation will favor those whose obstacles were completely alleviated and can effectively interact with the model. As a result, individuals are forced into areas (job selections, programs, schools, etc.) where they face fewer (no) access and utilization obstacles. Since some individuals are more likely than others to have more (fewer) obstacles when interacting with models in certain areas, individuals self-select into group roles. As a result, it exacerbates stereotypes, further marginalizes disadvantaged groups, and leads to very skewed distributions (see, for example, the effects of masculine defaults in schools and offices \cite{Cheryan2020MasculineDI} and how group roles shape stereotypes \cite{Koenig2014EvidenceFT}). We, therefore, implore decision-makers to pay close attention to the proxy gaps, obstacles individuals face to access and utilize the model, and ground truth curation. At the minimum, decision-makers should ask themselves, \textit{what kind of obstacles will people face to access and utilize my model? Who is likely to face more (less) obstacles? Am I using observations from evaluation and proxy features to curate new ground truth?}

\subsubsection{Why don't decision-makers simply use the intended model?}\label{sec:bounded_dm}
Some of the reasons why decision-makers use models far from what's on the ground include the following: bounded rationality which drives the notion of good enough decisions \citep{Simon1990}, lack of domain expertise, use of generic proprietary decision-making tools, over-reliance on skewed past experiences or biased view of the world \citep{pastfuture, decisiontovote,Stanovich2008OnTR, Bruin2007IndividualDI}, failure to consider the heterogeneity of environments, among others. Other times, it's inevitable to use an proxy model far from the intended model because decision-makers can only observe the intended features once decisions take form in the environment. Regardless of why there exists a discrepancy between the two models, our work draws decision-makers towards understanding this discrepancy, formulating equity, and curating ground truth with it in mind. The Equity Framework, through its checkpoints, would help steer decision-makers towards better and equitable decision-making. To gear towards equity, decision-makers should ask themselves questions like \textit{which label function and features should I use for my measurement?  If I used another label, would I still use the same features? Which one is closer to what's on the ground? How do access and utilization change? Who is likely to benefit (lose) the most if I chose this model over that one?}

\section{Equity Modelling}\label{sec:equity_score}

In a one-shot setting, the decision-maker makes a decision and then verifies the correctness of those decisions. In immediate evaluation models, the decision-maker immediately evaluates the correctness of their decision, for example, the prediction of a dog could be verified by crosschecking the prediction with the dog image. In delayed evaluation, the decision-maker waits until the decision takes form in the physical and social environment to verify their decision, for example, prediction of loan worthiness.

The Equity Framework draws attention to some of the unrealized consequences of delayed evaluation in which positively qualified individuals are evaluated by different models. It's crucial to assesses model utilization which evaluates whether an individual is a true or false positive, and effects of alleviation of access and utilization obstacles, and model outcomes. In section 3, we gave mathematical formulations of equal access, equal outcomes, and equal utilization. Algorithm \ref{alg:equity_score} outlines how decision-makers can evaluate how equitable their decision-making model is. Furthermore, to best help decision-makers plan for equitable decision-making, we provide summary of questions in table \ref{table:questions} that decision-makers should consider when creating an equitable delayed evaluation decision-making model.

\paragraph{Equity Scoring}

Algorithm \ref{alg:equity_score} shown in appendix \ref{app:scoring} shows steps a decision-maker interested in equity modeling can take. For example, if loan officers are interested in equitable modeling, instead of solely focusing on perfecting model outcomes, they must also take note of obstacles individuals face, their model proxy gaps, and give focus to optimizing model outcomes, access, and utilization.

Algorithm \ref{alg:equity_score}  works as follows: when the decision-maker decides on the problem to solve, they select probable models to use. Each model has a pool of features, access policies, and model functions. The decision-maker chooses the proxy model to use and checks if the model is accessible above a certain access threshold.
If satisfactorily accessible, then, given the selected features, the decision-maker selects/creates a model function to make predictions that achieve optimal model outcomes.  Then, the decision-maker evaluates the proxy model positive predictions to ensure optimal model utilization. In doing so, they ensure the presence of policies to alleviate utilization obstacles and that the intended model achieves model outcomes.  If the current model doesn't achieve utilization, the decision-maker repeats the process until they achieve a desirable equity score. 
A decision-maker can set several different stopping criteria based on their budget or preference. A perfect equitable model would be one with model access, $\Psi(P)=1$ , model outcomes, $\Omega(P)=0$ , and model utilization, $\zeta(P)=1$. 

In algorithm \ref{alg:equity_score}  decision-makers do not check for feature, label, and obstacle proxy gaps prior. However, to ease the equitable modelling, a decision-maker should ideally check for the proxy gaps before assessing the proxy model for access. If the proxy gaps are small, it means the proxy model is close or equivalent to the intended model that depicts the social and physical environment decisions take form. Therefore, the decision-maker wouldn't incur as much in finding the best intended model function and policy, since those used in proxy model might be applicable during evaluation.

\paragraph{Modelling Questions}
Decision-makers modeling delayed evaluation predictive problems, for example, admit/reject school admission, grant/deny bail, should at the very least ask themselves questions in table \ref{table:questions}.  The questions we provide in table \ref{table:questions} in appendix \ref{app:questions} are meant to guide decision-makers towards equitable modelling. We assume that when the decision-maker chooses features and labels to use for decision-making, some individuals will have obstacles accessing and utilizing the model. We also assume that there exists a set of policies that can alleviate obstacles the individuals face to access or utilize a given model.

\section{Case Study: Student Admission}\label{sec:case_study}
\subsection{Setting} We use the student admission case study to show the flow of the Equity Framework. We investigate access, outcome, and utilization when school admission decision-makers interested in determining who to admit use the ``admissibility'' proxy model instead of the ``likelihood to thrive'' intended model. The ``admissibility'' proxy model has an  ``admissible'' model function and uses features ``sex'', ``test\_scores'',  ``essay'', ``grades'', ``letter of recommendation'', ``extracurricular'' and applicant's target label, ``y''. On the other hand, the ``likelihood to thrive'' intended model has a ``performance'' model function and uses features ``sex'', ``health'', ``study time'', ``school absences'', ``travel time'', ``paid'', ``free time'',  ``romantic'', ``mother's education'',  ``father's education'', and the student's target label, ``y''.  

We curated this information from the UCI student performance dataset \citep{studentPerf}. Our determination of whether someone faced obstacles to access and utilize the model is subjective. We also assumed that while obstacle gap was zero, obstacles affected different features in the proxy and intended model.  To determine whether an individual faced obstacles, we based on the sum of their values for attributes ``paid tutoring'', ``family relationship'',  ``mother's job'',  ``father's job'', ``mother's education'' and  ``father's education''. In the proxy model, obstacles affected features ``test\_scores'',  ``essay'' and grades''. In the intended model, they affected ``health'', ``study time'', ``school absences'', and `free time''. For the models that alleviated utilization and access obstacles, we assumed that the policies alleviated all obstacles individuals faced.

\subsection{Results} 

\begin{figure}[!htp]
    \centering
    \includegraphics[width=0.55\linewidth]{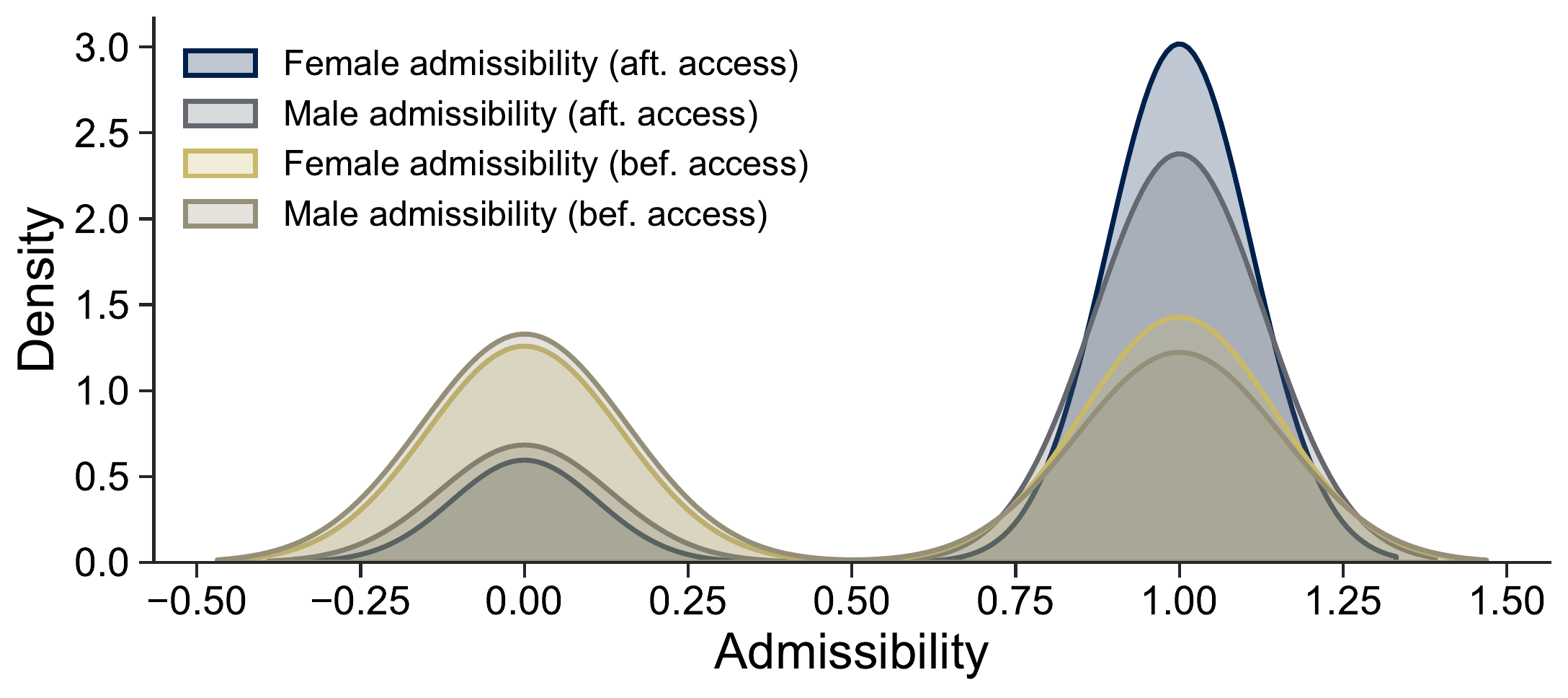}
    \caption{Equal access to the model increased admissibility for both the male and female populations, with a higher increase for the female group.}
    \label{fig:access_plot}
\end{figure}

\paragraph{Access} Our results show that in a model that ensures access $\Psi_{admissibility} = 1$, individuals' admissibility increases regardless of their protected group membership (see fig.\ref{fig:access_plot}).  
\raggedbottom

\paragraph{Outcomes} In addition, when the model checks for and ensures access and outcomes, it has has the lowest EO violation (see fig.\ref{fig:eo_plot}), and overall performs betters on all groups (see appendix fig.\ref{fig:outcome_plot}).  We also note that the unequal outcome and unequal access model had the worst performance on all groups. 
\begin{figure}[!htp]
    \centering
    \includegraphics[width=0.4\linewidth]{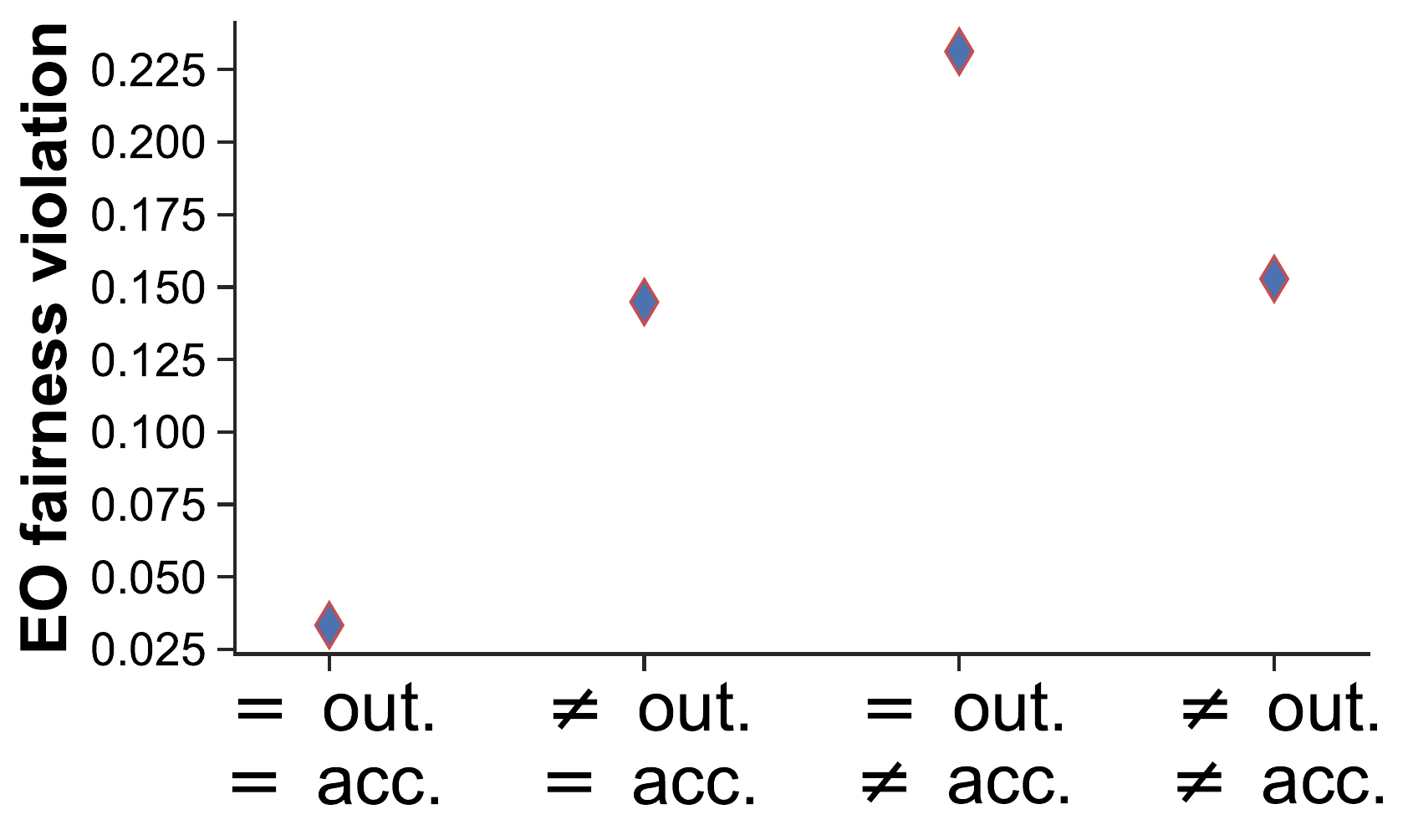}  
    \caption{An equal outcome and equal access model had the lowest EO violation and unequal access models had the highest EO violation on the male and female population}
    \label{fig:eo_plot}
\end{figure}
\raggedbottom
Performance results of (un)equal outcome and (un)equal access models (see appendix fig.\ref{fig:outcome_plot}) show that since model outcomes rely on the received data, it might be hard to notice the effect of access on model outcomes and performance of a sophisticated decision-making model. This blind spot highlights why decision-makers might ignore and sometimes forget the alleviation of obstacles and why it might seem like the model is fair when it's increasing the exclusion of marginalized groups. However, as fig.\ref{fig:util} shows, alleviation of access and utilization obstacles is crucial for equal utilization and future ground truth curation. 

\begin{figure}[!htp]
    \centering
    \includegraphics[width=0.87\linewidth]{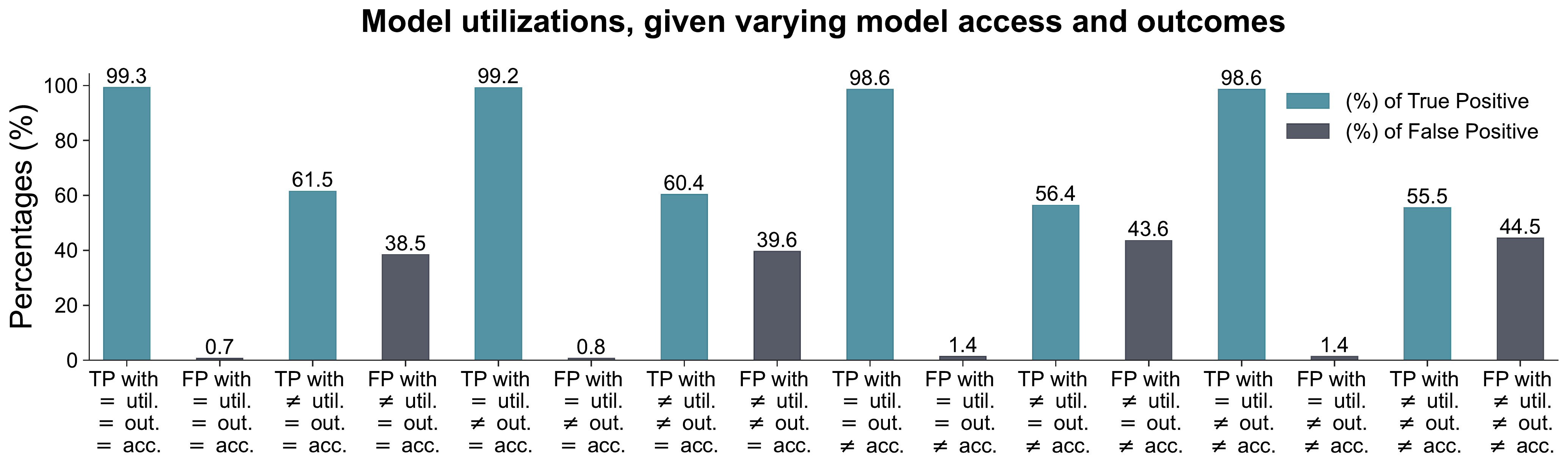}  
    \caption{The percentage of true positives decreased with decreased insurance of equal access, equal outcomes, and equal utilization, while the false positives increased.(meanings: $=$ is equal, $\neq$ is unequal, util. is utilization, out. is outcome, and acc. is access)}
    \label{fig:all_util}
\end{figure}
\raggedbottom

\paragraph{Utilization} When evaluating the admitted individuals for utilization, we found that the equal utilization, equal outcomes, and equal access model had the highest percentage of true positives ($99.3\%$) and the lowest percentage of false positives ($0.7\%$) (fig.\ref{fig:all_util}). The unequal utilization outcomes and access model had the lowest percentage of true positives ($55.5\%$) and the highest percentage of false positives ($45.5\%$).  In general, when compared to models with equal access, unequal access models irrespective of insurance of outcomes had lower true positives and higher false positives (less utilization) (fig.\ref{fig:all_util}).

Of the $0.7\%$ false positives in the equal access, equal outcomes, and equal utilization model, $100\%$ were male. Of the $44.5\%$ of false positives in the unequal access, unequal outcomes, and unequal utilization model, $41.5\%$ were male and $58.5\%$ were female.  Of the $39.6\%$ of false positives in the equal access, unequal outcomes, and unequal utilization model, $35.0\%$ were male and $65.0\%$ were female. In general, the female group had higher percentages of false positives than the male group since they had higher false positives in unequal utilization models, which generally had higher false positives than equal utilization models as seen in figure \ref{fig:util} in the appendix. Therefore, in general, ground truth curation (red lines in fig.\ref{fig:equity}) will favor the male group more than the female group.

Lastly, we provide an abstract pictorial example (see appendix \ref{appendix:pictorial}) to demonstrate a specific scenario in which our mathematical formulations are sufficiency, and aspects to look out for in equity modeling.

\section{Discussion, Limitations, and Conclusion}\label{sec:discussion}

In this work, we highlighted the main components ---equal model access, outcome, and utilization--- of achieving equitable decision-making. 
Our mathematical formulation and case study results showed how obstacles impede individuals from accessing and utilizing models and how alleviating these obstacles gives individuals a chance to interact with the models efficiently.  In addition, not all equal outcome models achieve equal access. The reason is, to compute model outcomes, decision-makers rely on the data received, and the data received is affected by variations of individuals' access to the model. Therefore, a model that neither checks nor ensures equal access can still have higher model outcomes.

We argued that since decision-makers are likely to choose models that deviate from those that depict the social and physical environment decisions are situated, not alleviating access obstacles is suboptimal. The reason is, while decision-makers might qualify obstacle-refrained individuals through changing the decision-making model, these obstacles resurface in utilization, causing individuals to appear as false positives. For example,  assume the proxy model falsely classifies positive or intentionally picks two individuals on the line: individual (a) advantaged and (b) disadvantaged. Individual (a) is more likely to be classified as a true positive because they have a higher chance of fully utilizing the model, while (b) a false positive.  The outcome would then confirm and or increase the biases of the decision-maker about the worthiness of the disadvantaged.  An unequal utilization model, therefore, leads to wrong, biased and incomplete ground truth. It's a cycle of inequalities feeding and reinforcing historical biases.

In addition, due to the ignorance of the gap between the proxy and intended model, blame is wrongly assigned, unfairness gets prolonged, and the welfare of the systems doesn't improve. The reason is it's harder to unravel, highlight and alleviate unknown obstacles, and measurements and curated ground truth are inaccurate and incomplete.  We, therefore, hope our work shifts the conversation on what fair predictive decision-making should look like and how to measure it.

While our formulations were more focused on strictly equal outcomes, equal access, and equal utilization, we realize that in practice, as depicted in our case study, this might be hard to achieve. This paper aims to, however, spark a conversation about what fairness should entail. That decision-makers try to ensure individuals have access to the model, outcomes from the models, and fully utilize the model. Future work could explore optimal relaxations of these formulations.

Although formally defining equitable decision-making is complex, our work attempts to draw concepts to make the definition more concrete and lay a foundation for defining equity in ML for predictive decision-making. To mathematically define obstacles and how they hinder access and utilization, we assumed that the effects of obstacles individuals face could only be seen directly in feature values.  
We see several crucial directions in obstacle alleviation to pursue further work. First, because of reliance on feature values to do obstacle alleviation, ensuring obstacle alleviation is robust to feature misspecification is important. Second, future work could investigate and model all the effects of obstacles because obstacles might directly affect both features and labels. Third, access to values of $\vec{x}$ and $\vec{z}$ might be heavily noisy which might lead to over or underestimation of the effects of obstacle alleviation. Mathematically formulating obstacle alleviation based on individual feature values is expensive and poses multiple complexities. However, group alleviation of obstacles raises intersectionality issues that are even harder to efficiently/correctly address. It is, therefore, important to carefully study a given problem in which equal access is to be achieved and choose the most applicable route of obstacle alleviation.

Our mathematical formulations of access, outcome and utilization are geared more towards the insurance of equity for the individuals decision-makers are classifying. 
Although these formulations could also benefit decision-makers, for example, low attrition rates, reduced rate of bad loans, the benefit might not be instant. In addition, to effectively achieve equity, decision-makers must exercise intentional thinking and seek the help of experts, for example, domain experts and legal consultants, to select the most appropriate features and suitable model function for a task, and discover obstacles people face and choose the best policies to alleviate those obstacles. We believe the process is an expensive, non-trivial, and altruistic endeavor. Therefore, to incentivize decision-makers towards equitable decision-making, new formulations that consider institutional utility and show a cost-benefit analysis could be helpful. In addition, we hope that a well-built equity assessment tool based on the Equity Framework would help decision-makers have checkpoints for what their model does or doesn't achieve en route to equitable decision-making.

\section*{Acknowledgements}
We thank Samni Koyejo for the immensely helpful feedback on the manuscript. Keziah Naggita was supported in part by the National Science Foundation under grant CCF-1815011 and by the Simons Foundation under the Simons Collaboration on the Theory of Algorithmic Fairness. Any opinions, findings, and conclusions or recommendations expressed in this material are those of the authors and do not necessarily reflect the views of the funding sources.

\bibliographystyle{unsrtnat}
\footnotesize{
\bibliography{refs}
}

\newpage
\appendix

\section{Equity Scoring Algorithm}\label{app:scoring}

\begin{algorithm}[htp!]
\caption{Determine the  model equity score}
\label{alg:equity_score}
\SetNoFillComment
\DontPrintSemicolon
\SetAlgoLined
\SetKwInOut{Input}{Input}
\SetKwInOut{Inputs}{}
\SetKwInOut{Output}{Output}
\SetKwInOut{Outputs}{}
\SetKw{Break}{break}
    \Input{$\{\mathcal{H}_{sP}, \mathcal{D}_{P}, \Phi_{sP}\}$ \tcp{proxy model space (model functions, features - labels distribution, \& policies)}}
    \Inputs{$\{\mathcal{H}_{sT}, \mathcal{D}_{T},  \Phi_{sT}\}$ \tcp{intended model space (model functions, features - labels distribution, \& policies)}}
    \Output{$score$}
    
    $\tau = 0.85$ \tcp{access and utilization threshold}
    $\tau_{o} = 0.15$ \tcp{outcomes threshold}
    \While{true}{
        $score \leftarrow 0$\;
        $\mathcal{H}_{P} \thicksim  \mathcal{H}_{sP}, (\mathcal{X}_{P}, \mathcal{Y}_{P}) \thicksim \mathcal{D}_{P}, (\mathcal{X}_{PT}, \mathcal{Y}) \thicksim \mathcal{D}_{P}, \mathcal{O}_{P}, \Phi_{P} \thicksim \Phi_{sP}$\; 
        $\mathcal{H}_{T} \thicksim  \mathcal{H}_{sT}, (\mathcal{X}_{T}, \mathcal{Y}_{T}) \thicksim \mathcal{D}_{T}, \mathcal{O}_{T}, \Phi_{T} \thicksim \Phi_{sT}$\; 
        \underline{\small{BEFORE PREDICTION}}\;
        Determine how accessible the model is, $\Psi(P)$, (see. sec.\ref{subsec:equal_access})\;
        Compute equity score, $score \leftarrow score + \Psi(P)$\;
        \tcc{ensure model access is beyond threshold}
        \If{$\Psi(P) \geq \tau$}{
            $outcome = \infty$\;
            \tcc{ensure model outcome is above the set threshold}
            \While{$outcome \geq \tau_{o}$}{
                \underline{\small{DURING PREDICTION}}\;
                Train the proxy model function, $h_{P} \in \mathcal{H}_{P}$  given the train set $ (\mathcal{X}_{P}, \mathcal{Y}_{P})$  which can be partitioned into training and cross validation sets \;
                Given the trained intended model function and $\mathcal{X}_{PT}$, compute the corresponding $\mathcal{Y}_{PT}$\;
                Determine model outcomes, $\Omega(P)$ (see. sec.\ref{subsec:equal_outcomes})\;
                $outcome \leftarrow \Omega(P)$\;
                \If{$outcome \leq \tau_{o}$}{ 
                    Update equity score, $score \leftarrow score + \Omega(P)$\;
                    \Break 
                }\Else{
                   Re-sample, $\mathcal{H}_{P}$ and retrain model function.
                }
            }
            $\mathcal{B} = \{i\} \ | \ y_{PT,i} \in \mathcal{Y}_{PT} = 1$\;
            $utilization = 0$ \tcp{initialize utilization score}
            \tcc{ensure model utilization is above the set threshold}
            \While{$utilization \leq \tau$}{
                \underline{\small{EVALUATION OF PREDICTION}}\;
                Train the intended model function $h_{T} \in \mathcal{H}_{T}$, given the train set $ (\mathcal{X}_{T}, \mathcal{Y}_{T})$ which can be partitioned into training and cross validation sets \;
                Individuals in $\mathcal{B}$ develop and reveal their features, $\mathcal{X}_{TT}$  to the intended model\;
                Given the trained intended model function and $\mathcal{X}_{TT}$, compute the corresponding $\mathcal{Y}_{TT}$\;
                Determine model utilization, $\zeta(P)$, given $\mathcal{Y}_{PT}$ and  $\mathcal{Y}_{TT}$ (see. sec.\ref{subsec:equal_util})\;
                $utilization \leftarrow \zeta(P)$\;
                \If{$utilization \geq \tau$}{ 
                    Update equity score, $score \leftarrow score + \zeta(P)$\;
                    \Break 
                }\Else{
                   Re-sample intended model features, policy and  model function, and retrain model function 
                }
            }
            \Break 
        }\Else{
            Re-sample the input features or find a better access policy
        }
    }
    \Return{$score$} \tcp{return equity score}
\end{algorithm}

\section{Equity Modelling Questions}\label{app:questions}

In this section, we provide a summary of questions to guide decision-makers towards equitable decision-making. 

\begin{table}[htp!]
\small
\centering
\begin{tabular}{ |s|p{10cm}|}
\hline
\multicolumn{2}{|c|}{Guiding questions for an equitable decision-making model}\\
\hline
\textbf{Task} & \textbf{Questions}\\
\hline
Selection of the proxy model & 1) What class of model functions would be the best to use, $\mathcal{H}_{P}$?\\
& 2) What features will be the most predictive?\\
& 3) Given a selection of features, $\mathcal{X}_{P}$, what access obstacles, $\mathcal{O}_{P}$, will individuals face to access this model?\\
& 4) Who is most likely to face the most (fewest) access obstacles?\\
& 5) Which policy can alleviate the obstacles individuals face? Do I have the policy,  $\Phi_{P}$, to alleviate the obstacles? What is the model access $\Psi_{P}$?\\
& 6) How well does the model function perform on individuals in different groups, $\Omega(P)$ ?\\
& 7) How would a change in features affect accuracy, obstacles faced, and model access $\Psi_{P}$ ?\\ 
& 8) How well does this model reflect the physical and social environment in which decisions take form?\\
& 9) Does the chosen model achieve equal access or optimal access threshold score?\\ 
\hline
Selection of evaluation model & 1) What class of evaluation model functions would be the best to use, $\mathcal{H}_{T}$?\\
& 2) What are features I am I using for evaluation? What is the feature proxy gap, $\Gamma_{X} (\mathcal{X}_{P}, \mathcal{X}_{T})$?\\
& 3) What is the label proxy gap, $\Gamma_{L}(h_{P}, h_{T})$?\\
& 4) Given a selection of evaluation features, $\mathcal{X}_{T}$, what utilization obstacles will individuals face to utilize the model?\\
& 5) What is the obstacle gap?\\
& 6) Who is most likely to face the most (fewest) utilization obstacles?\\
& 7) If I changed the features, would that increase (decrease) the accuracy of evaluation results and  increase (decrease) utilization obstacles faced?\\
& 8) Which policy can alleviate the utilization obstacles individuals face? Do I have the policy,  $\Phi_{T}$, to alleviate the utilization obstacles?\\
& 9) How well does the evaluation model function perform on individuals in different groups?\\
& 10) Does the chosen model achieve equal utilization or optimal utilization threshold score?\\ 
\hline
Curation of ground truth & 1) Given the obstacle gap, label proxy gap, $\Gamma_{L}(h_{P}, h_{T})$, and feature proxy gap, $\Gamma_{X} (\mathcal{X}_{P}, \mathcal{X}_{T})$, should I use proxy or evaluation model features/labels or both?\\
& 2) If I choose these features/labels, given utilization, $\zeta(P)$, access, $\Psi(P)$, and outcome, $\Omega(P)$, who is most likely to be misrepresented in the new ground truth? Do I exhaustively capture obstacles individuals face?\\
\hline
\end{tabular}
\vspace{3mm}
\caption{\label{table:questions}Questions decision-makers should at the minimum ask themselves}
\end{table}

\section{Case Study Results}\label{appendix:case-study}

\begin{figure}[!htp]
\centering
\includegraphics[width=\linewidth]{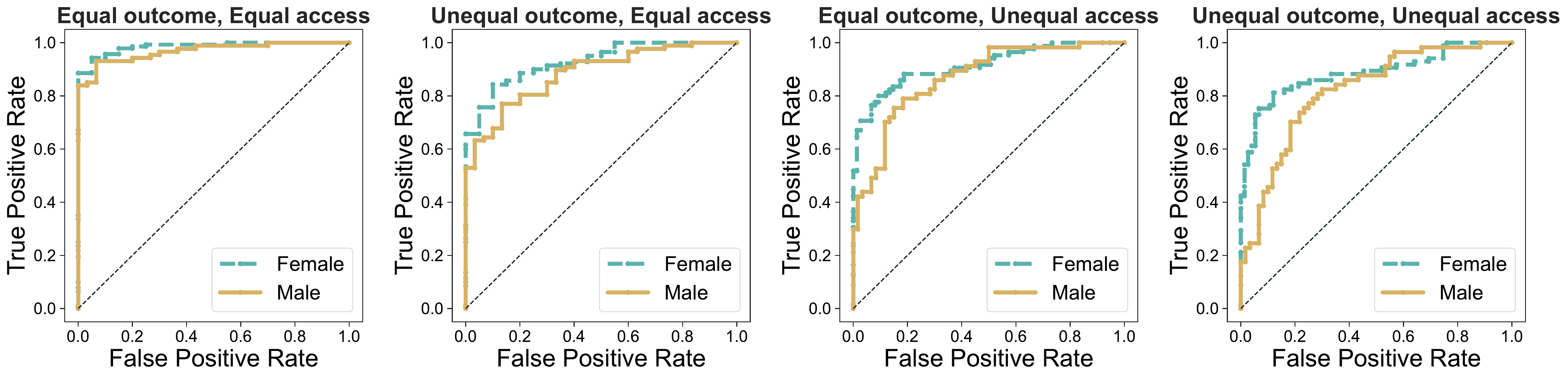}\label{fig:eo_violation}
\caption{An equal outcome and equal access model had the best performance on both male and female individuals, and unequal access models had the worst performance on the male and female population. In general, all models performed comparably better on the female group than the male group. (meanings: $=$ is equal, $\neq$ is unequal, out. is outcome, and acc. is access)}
\label{fig:outcome_plot}
\end{figure}
\raggedbottom

\begin{figure}[!htp]
\centering
\subfloat[]{\includegraphics[width=0.87\linewidth]{arXiv-figs/util.pdf}\label{fig:all_utils}}
\hspace{0.1\textwidth}
\subfloat[]{\includegraphics[width=0.87\linewidth]{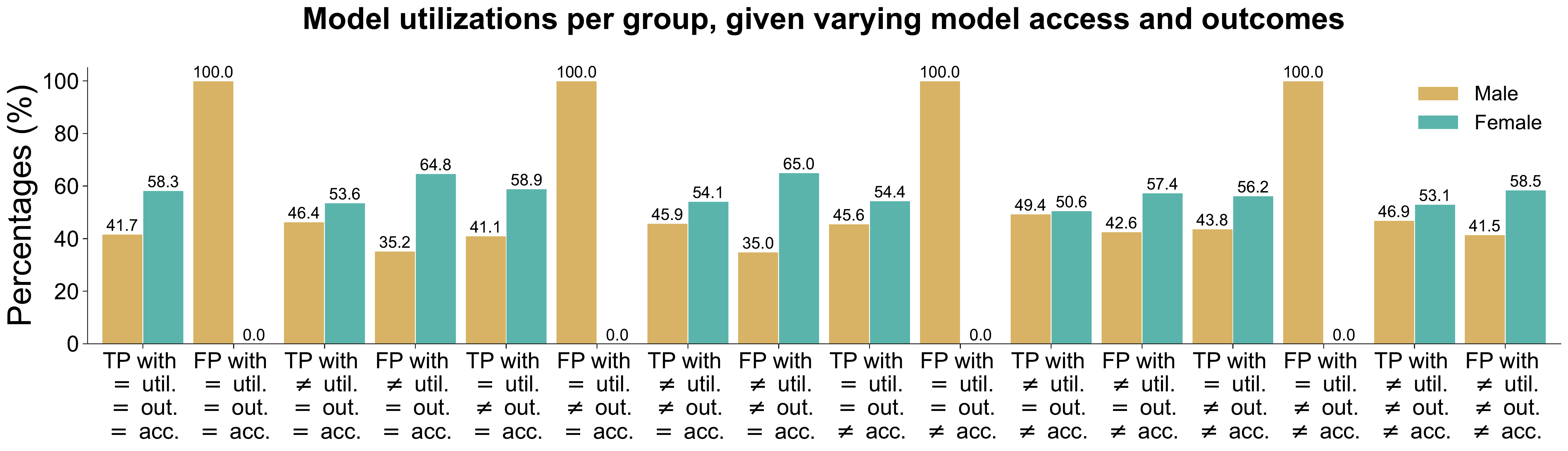}\label{fig:group_util}}
\caption{In (a), the percentage of true positives decreased with decreased insurance of equal access, equal outcomes, and equal utilization, while the false positives increased. In (b), the female group had consistently had higher false positives than males in the unequal utilization models, and the male group had higher false positives in the equal utilization models. (meanings: $=$ is equal, $\neq$ is unequal, util. is utilization, out. is outcome, and acc. is access)}
\label{fig:util}
\end{figure}
\raggedbottom

\section{Pictorial Example}\label{appendix:pictorial}
In this section we provide a specific abstract example and settings where specifications of our mathematical formulations might be sufficient. 

\begin{figure}[!htp]
    \centering
    \includegraphics[width=\linewidth]{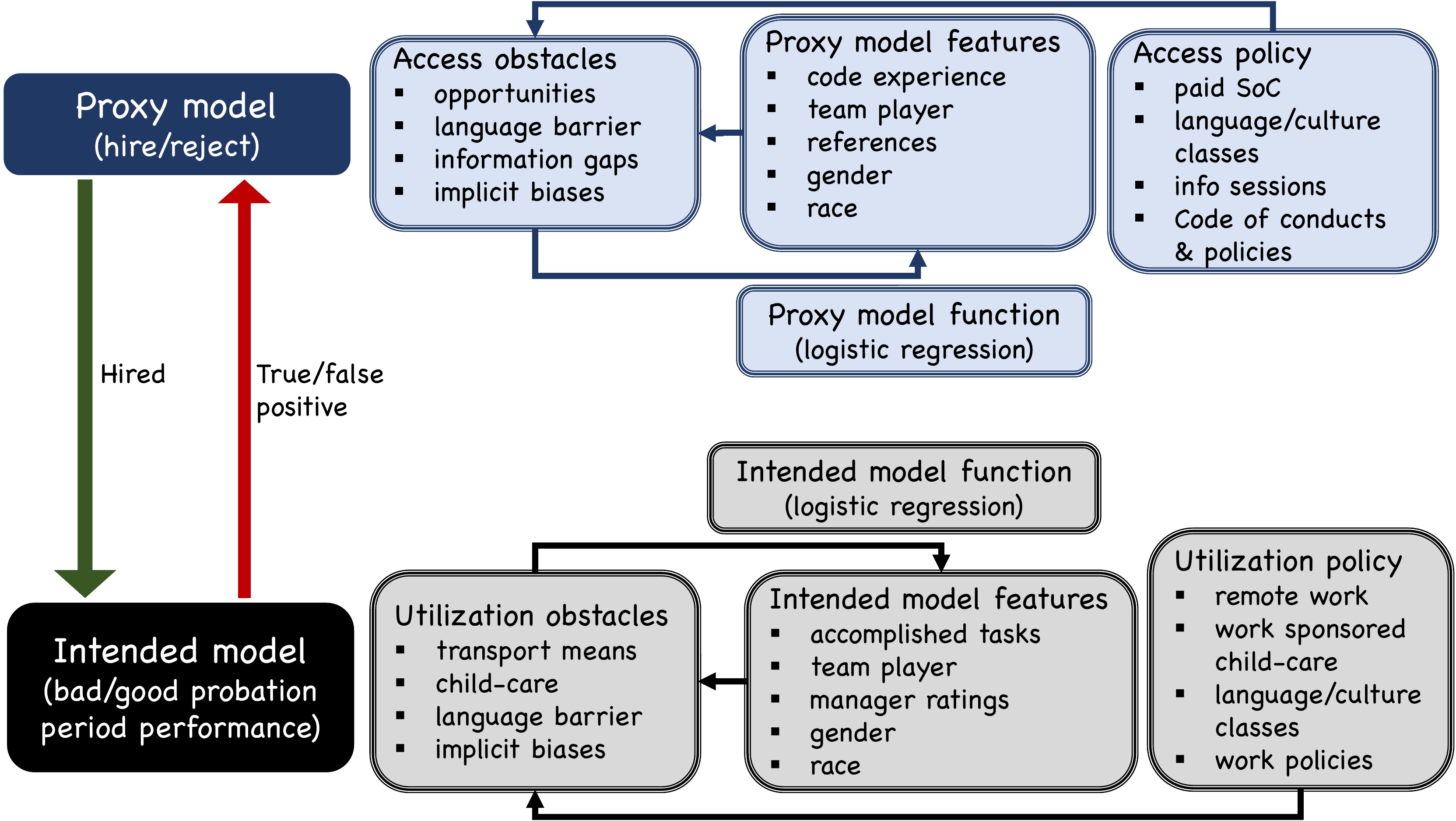}  
    \caption{A simplistic example of a junior software developer hiring model}
    \label{fig:abstract_example}
\end{figure}
\raggedbottom

Figure \ref{fig:abstract_example} illustrates a simplistic junior software developer hiring model. The decision-maker uses the hiring proxy model to determine who to hire and uses the probation period performance to evaluate their prediction. In the proxy model, if an applicant had more paid opportunities to exercise their coding skills, for example, paid summer of code (SoC), their coding experience score would be higher. If an individual faced language barrier, they would have a lower team player score. In this example, alleviation of obstacles leads to $\vec{z}$ dominating $x$ as described in section \ref{subsec:equal_access}.

Let the feature importance scores of the proxy model function (logistic regression) be as follows; ``code experience'' - $0.5$, ``team player'' - $0.2$, ``references'' - $0.2$, ``gender'' - $0.05$, and ``race'' - $0.05$'. Let the features scores of the proxy model function (logistic regression) be as follows; ``accomplished tasks'' - $0.5$, ``team player'' - $0.2$, ``manager ratings'' - $0.2$, ``gender'' - $0.05$, and ``race'' - $0.05$'. In this setting, we can easily compute the feature proxy gap would be $(1,0,1,0,0)$ and the label proxy gap $(0.5,0.0,0.3,0.0,0.0)$ using formulations given in section \ref{subsec:equal_util}.  
Based on this setting, we can compute access, outcome and utilization using the definitions given in section \ref{sec:framework}. The red line in figure \ref{fig:abstract_example} is synonymous to the red line in figure \ref{fig:equity}. It indicates how decision-makers who might be oblivious to the intended model might use the observations $\mathcal{Y}_{TT}$ minus how they came to be to curate ground truth.

\paragraph{Comments on the setting}
First, note that the list of features is not necessarily exhaustive. For example, we did not include social-economic status in the proxy model features. However, one's social-economic status could affect one's code experience dating to having a laptop/internet and or time to build up their coding experience. It is, therefore, crucial to ensure the decision-making involves several experts to ensure selection of appropriate features and consequently proper identification of obstacles.

Lastly, the obstacles we included in this example model setting are not exhaustive and may not apply to all individuals. Again, it is crucial to involve other experts, for example, domain experts, when designing models. The experts would help decision-makers identify the obstacles individuals face to access/utilize models and choose the best policies to alleviate them. Additionally, decision-makers should avoid making assumptions about the existence or absence of obstacles.

\end{document}